\documentclass[12pt]{article}
\usepackage{epsfig}

\def\lsim{\raise0.3ex\hbox{$<$\kern-0.75em\raise-1.1ex\hbox{$\sim$}}}
\def\gsim{\raise0.3ex\hbox{$>$\kern-0.75em\raise-1.1ex\hbox{$\sim$}}}

\def\beq{\begin{equation}}
\def\eeq{\end{equation}}
\def\bea{\begin{eqnarray}}
\def\eea{\end{eqnarray}}
\def\noi{\noindent}
\begin{document}

\rightline{CERN-TH/2003-066}
\rightline{LPT Orsay 03-21}
\rightline{UCOFIS 2/03}
\rightline{April 2003}

\vspace{0.5cm}

\begin{center}
{\Large\bf Nuclear Structure Functions}\\ 
{\Large\bf at Small $x$ from}\\
{\Large\bf Inelastic Shadowing and Diffraction}
\vspace{1cm}
 
N. Armesto$^{a,1}$, A. Capella$^{b,2}$,
A. B. Kaidalov$^{c,3}$,\\
J. L\'opez-Albacete$^{a,d,4}$ and C. A. Salgado$^{a,5}$
 
\vspace{0.2cm}
 
{\it $^a$ Theory Division, CERN, CH-1211 Gen\`eve 23, Switzerland
}\\
{\it $^b$ Laboratoire de Physique Th\'eorique, Universit\'e de Paris XI,
B$\hat{a}$timent 210,}\\
{\it F-91405 Orsay Cedex, France 
} \\
{\it $^c$ Institute of Theoretical and Experimental Physics, B. Cheremushkinskaya
25, Moscow 117259, Russia
} \\
{\it $^d$ Departamento de F\'{\i}sica, M\'odulo C2, Planta baja,
Campus de Rabanales, Universidad de C\'ordoba, 14071 C\'ordoba, Spain
} \\
 
\end{center}

\vspace{1.cm}
{\small
Nuclear structure functions at
small $x$ and small or moderate $Q^2$
are studied using the relation with diffraction
on nucleons which arises from Gribov's Reggeon Calculus. A reasonable
description of experimental data is obtained with no fitted parameters.
A comparison with other
models and
predictions for future lepton-ion colliders
are provided. Consequences
for the reduction of multiplicities in nucleus-nucleus collisions at
energies of RHIC and LHC are examined.
}

\vfill


\noindent{\small $^1$ E-mail: Nestor.Armesto.Perez@cern.ch}\\
\noindent{\small $^2$ E-mail: Alphonse.Capella@th.u-psud.fr}\\
\noindent{\small $^3$ E-mail: kaidalov@heron.itep.ru}\\
\noindent{\small $^4$ E-mail: Javier.Lopez.Albacete@cern.ch}\\
\noindent{\small $^5$ E-mail: Carlos.Salgado@cern.ch}

\newpage

\section{Introduction} \label{intro}

The study of nuclear structure functions has become a fashionable subject.
Apart from its intrinsic interest,
such analysis has a great impact on the interpretation of
results from heavy ion experiment.
At small values of the
Bjorken variable $x$ ($\lsim \ 0.01$,
shadowing region), the
structure function $F_2$ per nucleon turns out to be smaller in nuclei than
in a free nucleon \cite{Arneodo:1992wf,Geesaman:1995yd}. Several explanations
to this shadowing have been proposed.

On the one hand, some models use the
fact that, in the rest frame of the nucleus
the incoming photon splits into a $q\bar q$ pair long
before reaching the nucleus, and this $q\bar q$ pair interacts with it
with typical hadronic cross sections, which results in absorption
\cite{brodsky,nikolaev,nikolaev2,kope,noso,meu,kope2}.
Thus nuclear shadowing is a
consequence of
multiple scattering which in turn
is related to diffraction \cite{kope,funnuc,fgms}. This relationship
will be developed in this
paper.
Equivalently, in a
frame in which the nucleus is moving fast, gluon recombination due
to the overlap of the gluon clouds from different nucleons reduces the
gluon density
in the nucleus
\cite{glr,mq}.
These studies have received great theoretical impulse with the development of
semiclassical ideas in QCD and the appearance of non-linear equations for
evolution in $x$ in this
framework
(see \cite{McLerran:1993ni,Mueller:2002kw,Venugopalan:1999wu,cargese}
and references therein;
also
\cite{carlinhos} for a simple geometrical approach in this framework).

On the other hand, other approaches \cite{gnucl1,gnucl2,gnucl3}
do not address the origin of shadowing but only its evolution with $\ln Q^2$:
parton
densities inside the nucleus are parametrized at some scale
$Q_0^2$ and then evolved using the DGLAP \cite{dglap}
evolution equations.

The results from different models usually depend on phenomenological
assumptions and their predictions (notably for small values
of $x$
which are of uttermost importance to compute particle production at RHIC and
LHC) turn out to be very different. For example, concerning
the $Q^2$-dependence of shadowing, it can be either constant
\cite{nikolaev,nikolaev2,kope,noso,meu,kope2},
or die out logarithmically
\cite{gnucl1,gnucl2,gnucl3}
or behave as a higher-twist \cite{glr,mq}.

In this paper we will use the relation of diffraction to nuclear
shadowing which arises from Gribov Theory \cite{gt}, Reggeon Calculus \cite{rc}
and the AGK rules \cite{agk}. In this way we obtain a
parameter-free description of nuclear structure functions in the shadowing
region valid for $x<0.01$ and $Q^2<10$ GeV$^2$,
using a model for $F_2$ and $F_{2\mathcal{D}}$ \cite{model1,model2}.
The same strategy has been used in \cite{funnuc,fgms}, but our extrapolation
to smaller $x$ or higher $W^2$ is more reliable than that of \cite{funnuc} due
to the model employed for the nucleon; besides, our description is valid for
small $Q^2$ while that of \cite{fgms} applies to $Q^2\geq 4$ GeV$^2$.
In
Section \ref{model}
the model will be described. In Section \ref{numer} numerical results will
be presented together with comparisons with experimental data and with other
models.
In Section \ref{mulred}
the model will be applied to calculate the multiplicity reduction
factors \cite{hip,aldol} relevant to compute particle production in heavy
ion collisions at RHIC and LHC. Finally, the last Section will contain our
conclusions.

\section{Description of the model} \label{model}

We assume that nuclei are made of nucleons in the spirit of the Glauber model.
In order to relate diffraction on nucleons with nuclear shadowing, we will
follow the procedure
explained in \cite{funnuc}. The $\gamma^*$-nucleus cross
section can be expanded in a multiple scattering series containing
the
contribution from 1, 2,$\dots$ scatterings:
\beq
\label{eq1}
\sigma_A = \sigma_A^{(1)}+ \sigma_A^{(2)}+\cdots.
\eeq
$\sigma_A^{(1)}$ is simply
equal to $A\sigma_{{\rm nucleon}}$. Let us consider now the
first correction to the non-additivity of cross sections which comes from
the second-order rescattering $\sigma_A^{(2)}$.
In Fig. 1 diffractive DIS is shown in both the infinite momentum frame
and in the rest frame of the nucleon. In Fig. 2 it becomes clear that the
square of such contribution is equivalent to a double exchange with a cut
between the exchanged amplitudes, a so-called diffractive cut.
To compute the first
contribution to nuclear shadowing $\sigma_A^{(2)}$, which comes from these
two exchanges, we need
its total contribution to the $\gamma^*$-nucleon cross section,
which arises from cutting the two-exchange amplitude in
all possible ways (between the amplitudes and the amplitudes themselves in
all possible manners). For purely imaginary amplitudes,
it can be shown \cite{rc,agk} that this
total contribution
is identical to minus the contribution from the diffractive cut. Thus
diffractive DIS is directly related to the first contribution
to nuclear shadowing.
The final expression reads
\beq
\label{eq2}
\sigma_A^{(2)}=-4\pi A(A-1)\int d^2b\ T_A^2(b)
\int _{M^2_{min}}^{M^2_{max}}dM^2 \left.
\frac{d\sigma^{\mathcal{D}}_{\gamma^*{\rm p}}}{dM^2dt}\right\vert_{t=0}
F_A^2(t_{min}),
\eeq
with $T_A(b)=\int_{-\infty}^{+\infty} dz\rho_A(\vec{b},z)$ the nuclear profile
function normalized to 1, $ \int d^2b\ T_A(\vec{b})=1$, and
$M^2$ the mass of the diffractively produced system.
The usual variables for diffractive DIS: $Q^2$, $x$, $M^2$ and $t$,
or
$x_P=x/\beta$, $\beta=\frac{Q^2}{Q^2+M^2}$, are shown in Fig. 1.

Coherence effects, i.e. the coherence length of the $q\bar q$ fluctuation of
the incoming virtual photon, is taken into account through
\beq
F_A(t_{min})=\int d^2b\ J_0(b\sqrt{-t_{min}})T_A(b),
\label{eq2-1}
\eeq
with $t_{min}=-m_N^2x_P^2$ and $m_N$ the nucleon mass.
This function is equal to 1 at $x\to 0$ and
decreases
with increasing $x$ due to the loss of coherence for $x>x_{crit}\sim
(m_NR_A)^{-1}$.

Let us briefly examine (\ref{eq2}). Here the real part of the Pomeron amplitude,
which is small for the value of the intercept which will be
used \cite{model2},
$\Delta=\alpha_P(t=0)-1=0.2$, has not been taken into account.
Also it has been deduced under the
approximation $R^2_A\gg R^2_N$, so the $t$-dependence of the $\gamma^*$-nucleon
cross section has been neglected.

For
$A>20$ a nuclear density in the form of a 3-parameter Fermi
distribution with parameters taken from \cite{DeJager:1974dg} will be employed
to compute both $T_A(b)$ and (\ref{eq2-1}).
For $2<A\leq 20$ 
a Gaussian profile function is used \cite{preston}:
\beq
\label{eq7}
T_A(b)=\frac{3}{2\pi R_A^2}\ \exp\left(-\frac{3b^2}{2R_A^2}\right),\ \ 
R_A=0.82A^{1/3}+0.58\ \ {\rm fm},
\eeq
but, in order to take into account the $t$-dependence for these
light nuclei, we make in the computation of form factors (\ref{eq2-1})
the substitution
\beq
\label{eq7-1}
R_A^2 \longrightarrow R_A^2+R_N^2,\ \ R_N=0.8 \ \ {\rm fm}.
\eeq
Finally, for deuteron the double rescattering contribution
has the form
\beq
\label{eq6}
\sigma_A^{(2)}=-2\int _{-\infty}^{t_{min}}dt  \int _{M^2_{min}}^{M^2_{max}}dM^2
 \left.\frac{d\sigma^{\mathcal{D}}_{\gamma^* {\rm nucleon}}}
{dM^2dt}\right\vert_{t=0}F_D
(t),
\eeq
where $F_D(t)=e^{at}$, $a=40$ GeV$^{-2}$.

The lower integration limit in (\ref{eq2}) and (\ref{eq6}) is $M^2_{min}=
4m_\pi^2
=0.08$ GeV$^2$,  while the upper one is taken from the condition:
\beq
\label{eq5}
x_P=x\left(\frac{M^2+Q^2}{Q^2}\right)\leq x_{Pmax}
\Longrightarrow M^2_{max}= Q^2\left(\frac{x_{Pmax}}{x}-1\right),
\eeq
with $x_{Pmax}=0.1$; this value was used in \cite{model2} motivated by the fact
that the model is only valid for $M^2\ll W^2$ or $x_P\ll 1$, i.e. a large
rapidity gap is required. In our
case, variations of $x_{Pmax}$ by a factor 2 do not affect the description
of nuclear shadowing at $x<0.01$ but the choice $x_{Pmax}=0.1$ is convenient as
it guarantees the disappearance of nuclear shadowing at $x\sim 0.1$ (see
below) as in the experimental data.

The relation between $\left.\frac{d\sigma^{\mathcal{D}}_{\gamma^*{\rm p}}}
{dM^2dt}\right\vert_{t=0}$ and $x_PF^{(3)}_{2\mathcal{D}}(Q^2,x_P,\beta)$ is
provided by the model \cite{model2}:
\bea
&x_PF^{(3)}_{2\mathcal{D}}(Q^2,x_P,\beta)&=
x_P\frac{Q^2}{4\pi^2\alpha_{em}}\int_{-\infty}^0 dt
\frac{d\sigma^\mathcal{D}_{\gamma^*{\rm p}} (Q^2,x_P,\beta,t)}{dx_Pdt}
\nonumber \\
&\Longrightarrow&
\left.\frac{d\sigma^\mathcal{D}_{\gamma^*{\rm p}} (Q^2,x_P,\beta)}{dM^2dt}
\right\vert_{t=0}=
\frac{4\pi^2\alpha_{em}B}{Q^2(Q^2+M^2)}x_PF^{(3)}_{2\mathcal{D}}(Q^2,x_P,\beta),
\label{eq3}
\eea
where the usual factorization has been assumed:
\beq
\label{eq4}
\frac{d\sigma^{\mathcal{D}}_{\gamma^*{\rm p}}(x,Q^2,M^2,t)}{dM^2dt}=
\left.\frac{d\sigma^{\mathcal{D}}_{\gamma^*{\rm p}}(x,Q^2,M^2)}{dM^2dt}\right|
_{t=0}e^{Bt},
\eeq
with $B=6$ GeV$^{-2}$ (as in \cite{gbw}, see the discussion there; this value
is
slightly smaller than the experimental values
$7.2\pm 1.1 ({\rm stat.})^{+0.7}_{-0.9}({\rm syst.})$ GeV$^{-2}$ \cite{zeus}
at $\langle Q^2\rangle =8$ GeV$^2$ and $6.8\pm 0.9 ({\rm stat.})^{+1.2}
_{-1.1}({\rm syst.})$ GeV$^{-2}$ \cite{zeus2} for photoproduction).
Note that $\left.\frac{d\sigma^{\mathcal{D}}_{\gamma^*{\rm p}}(x,Q^2,M^2)}
{dM^2dt}\right|
_{t=0}$ can be obtained directly from $\sigma_{tot}$.
However, the model for diffraction we are using \cite{model2}
has mainly been tested 
after integration in $t$ (most available data are integrated in $t$). For this
reason, we use the integrated expression together with the experimental
value of $B$. While this is legitimate at present values of $x$, it can lead
to an underestimation of shadowing at very small $x$, due to the increase
of $B$ with energy\footnote{Nevertheless, the effect is not too large:
we have checked that an increase 
of $B$ from 6 to 7.2 GeV$^{-2}$ produces an increase of shadowing for Pb
of at most
10 \% at $x=10^{-7}$. As estimates indicate an increase $\lsim 50$ \% in $B$
for the smallest $x$ we have studied, $x=10^{-7}$, the increase of shadowing
due to this
effect would be at most $\sim 25$ \% for these values of $x$.}.



The model in \cite{model2} is based on the dipole picture of the photon and
contains two components. The small-distance ($S$) component corresponds to
transverse distances $r$ between the $q$ and the $\bar q$ of the dipole
such that $r<r_0$,
and the large-distance ($L$) component to $r>r_0$, with $r_0=0.2$
fm.
In each component a quasi-eikonal iteration
is introduced
in order to enforce unitarity. Reggeon and
Pomeron exchanges
are allowed. For diffraction, a third component is used, namely a
contribution from the triple
interaction of Reggeons and Pomerons. This model has
been designed to describe the small $x<10^{-2}$, small or moderate $Q^2<10$
GeV$^2$ region, and it contains the basic ingredients which allow its safe
extrapolation\footnote{In order to use the model for
larger $x$, $0.01<x<0.1$, we have made some modifications in \cite{model2}:
there, in
Eq. (26) $\beta_{min}$ in the normalization denominators has been set to 0, and
in Eq. (25) the Reggeon-Reggeon contribution has been ignored. These two
changes slightly modify the description of diffraction but we have
checked that the agreement with experimental data is as good as in the
original version of the model.}
to very small $x$ or high $W^2$.

Eq. (\ref{eq2}) corresponds to the case with only two scatterings. Its
extension to include higher order rescatterings is model-dependent. We will
use
two models:
a Schwimmer unitarization \cite{schwimmer} which is obtained from a summation
of fan diagrams with triple Pomeron interactions,
\beq
\label{eq8}
\sigma^{Sch}_{\gamma^*A}=\sigma_{\gamma^*{\rm nucleon}}\int d^2b
\ \frac{AT_A(b)}{1+(A-1)f(x,Q^2)T_A(b)}\ ,
\eeq
and an eikonal unitarization,
\beq
\label{eq9}
\sigma^{eik}_{\gamma^*A}=\sigma_{\gamma^*{\rm nucleon}} \int d^2b
\ \frac{A}{2(A-1)f(x,Q^2)}\left\{1-\exp{\left[-2(A-1)T_A(b)f(x,Q^2)\right]
}\right\},
\eeq
where we use the relation
$\sigma_{\gamma^*{\rm nucleon}}=\frac{4\pi^2\alpha_{em}}{Q^2}F_2(x,Q^2)$
valid at small $x$. Here, $F_2(x,Q^2)$ is the nucleon structure function,
taken from \cite{model2}.
Both expressions (\ref{eq8}) and (\ref{eq9}),
expanded to the first non-trivial order,
reproduce the second rescattering result (\ref{eq2}). Eikonal unitarization will
produce larger shadowing than Schwimmer, as can be expected by comparing the
second non-trivial order in the expansion of both expressions.
Finally,
\beq
\label{eq11}
f(x,Q^2)=\frac{4\pi}{\sigma_{\gamma^*{\rm nucleon}}}
\int _{M^2_{min}}^{M^2_{max}}dM^2 \left.\frac{d\sigma^{\mathcal{D}}
_{\gamma^*{\rm p}}}{dM^2dt}\right\vert_{t=0}F_A^2(t_{min})
\eeq
as required by consistency with (\ref{eq2}).

The shadowing in nuclei is usually studied through the
ratios of cross sections per
nucleon for
different nuclei, defined as
\beq
\label{eq12}
R(A/B)=\frac{B}{A}\frac{\sigma_{\gamma^{*}A}}{\sigma_{\gamma^{*}B}}\ .
\eeq
In the simplest case of the ratio over nucleon (equivalent to proton at small
$x$ where the valence contribution can be neglected), we get:
\beq
\label{eq13}
R^{Sch}(A/{\rm nucleon})=\int d^2b\ \frac{T_A(b)}{1+(A-1)f(x,Q^2)T_A(b)}\ ,
\eeq
\beq
\label{eq14}
R^{eik}(A/{\rm nucleon})=\int d^2b
\ \frac{1}{2(A-1)f(x,Q^2)}\left\{1-\exp{\left[-2(A-1)T_A(b)f(x,Q^2)
\right]}\right\}.
\eeq
To calculate shadowing in photoproduction,
$x$ is no longer a relevant kinematical variable.
Instead we use the $\gamma^*$-nucleon center of mass energy $W^2$.

In our framework shadowing can also be studied
as a function of the impact parameter $b$:
\beq
\label{eq15}
R(A/{\rm nucleon})^{Sch}(b)=\frac{1}{1+(A-1)f(x,Q^2)T_A(b)}\ ,
\eeq
\beq
\label{eq16}
R(A/{\rm nucleon})^{eik}(b)=\frac{1}{2(A-1)T_A(b)f(x,Q^2)}
\left\{1-\exp{\left[-2(A-1)T_A(b)f(x,Q^2)\right]}\right\}.
\eeq

Finally, the region of applicability of our model is the same as
that of the model for
diffraction on nucleon
\cite{model2}, i.e. small $x\lsim 0.01$ and small or moderate
$Q^2 \lsim 10$ GeV$^2$, including photoproduction.

\section{Numerical results} \label{numer}

In our model and in \cite{model2} we work in the small $x$ region
and thus no distinction is made between protons and neutrons.
Although usually
joined with straight
lines, our results are computed at the same $\langle x\rangle$
and $\langle Q^2\rangle$ as the experimental data. For the latter, inner error
bars show statistical errors, and outer error bars correspond to statistical and
systematical errors added in quadrature.

In Figs. 3-6 a comparison with experimental data at small $x$
from E665 \cite{e665-2,e665-1}
and NMC \cite{nmc-1,nmc-3,nmc-2} is presented. As
expected, eikonal unitarization produces larger shadowing than Schwimmer.
The agreement with experimental data
is quite reasonable
taking into account that no parameters have been fitted to reproduce the data.
Two comments are in order: First, for C/D and Ca/D
in Fig. 3 which shows the
comparison with E665 data, shadowing
looks overestimated for $x\sim 0.01$, while in Fig. 5 which shows
the comparison with
NMC data, it looks underestimated. This corresponds to the known difference 
between
the results of both experiments for ratios over D, while the compatibility
is restored \cite{nmc-1} when ratios are computed over C.
Second, from Fig. 6 it becomes clear that the evolution with $Q^2$
in the model
is too slow at $x\sim 0.01$, a problem related with the lack
of DGLAP evolution in the model \cite{model2} (see
\cite{funnuc,fgms,novonoso}
for an application of DGLAP evolution to initial conditions).

In Fig. 7 a comparison of the results of our model with those of others is
shown, for $Q^2=3$ GeV$^2$ (except the results of \cite{fgms} which are at
$Q^2=4$ GeV$^2$). It can be seen that the results of different models
agree within 15 \%
at $x\sim 0.01$ where experimental data exist, while they differ
up to a factor 0.6 at $x=10^{-5}$. At this $x$, our results are the lowest ones
but
roughly agree with those of \cite{gnucl1} and with one set of \cite{fgms},
while the results from \cite{gnucl3} are the highest ones, and those
of \cite{meu,ina,jochen} and the second set of \cite{fgms} lie
in between.
Let us briefly comment on these models: In \cite{gnucl1,gnucl3}
an initial condition is parameterized at some $Q_0^2$ and then evolved
using DGLAP; the initial condition is fitted
from the comparison of the evolved results with experimental
data (see \cite{novocarlos} for a comparison between these two models).
\cite{meu} is a model which uses a saturating ansatz for the
total $\gamma^*$-nucleon cross section in the proton, which is introduced
in a Glauber expression for its extension to the nuclear case.
In \cite{fgms} some parameterization of hard
diffraction at $Q_0^2$, which as in the present work gives nuclear shadowing
through Gribov's
Reggeon Calculus, is employed;
this nuclear shadowing computed at $Q_0^2$ is used as initial condition for
DGLAP evolution.
In \cite{ina} a Glauber
ansatz provides with the initial condition for DGLAP evolution.
Finally, in \cite{jochen} a non-linear equation for small $x$ evolution
is numerically solved \cite{Gotsman:2002yy} and used in the nuclear case.
In view of the differences at small $x$ among different models, a measurement
of $F_{2}$ in nuclei with $\sim 10$ \% precision would be a sensitive test to
discriminate among them.
Lepton-ion colliders \cite{eic} could provide with such data.

In Fig. 8 our predictions for the ratios D, He, Li, C, Ca, Sn and Pb over
nucleon for $Q^2=0.5$, 2 and  5 GeV$^2$ are given for $x>10^{-8}$. Let us notice
that our model is designed for the small $x$ region and that no antishadowing
or any other effects relevant for $x\gsim 0.1$ have been introduced. The
disappearance of shadowing at $x\sim 0.1$ is a consequence of both the
coherence effects in (\ref{eq2-1}) and the vanishing integration domain in
(\ref{eq2}), see (\ref{eq5}).
In Fig. 9 results in photoproduction
for the same ratios as in Fig. 8
are given for
$W^2<10^5$ GeV$^2$, together with predictions for the evolution
of the ratios C and Pb over nucleon with impact parameter $b$. Values as low as
0.3 are reached for central Pb/nucleon. This evolution with centrality is
very important to compute the corresponding evolution of particle production
in nuclear collisions, and could also be measured in lepton-ion colliders
\cite{eic}.

As a last comment in this Section, let us discuss about
the twist structure of the model (i.e. its structure in powers of $1/Q^2$).
In the model of \cite{model2} the unitarity corrections to the 
$L$-component are all of order $1/Q^2$. On the contrary,
in the $S$-component the unitarity corrections are higher-twist 
(they can be expanded as a sum of terms, each one containing
an additional factor $1/Q^2$ as compared to
the previous one).
The fact that diffraction is related to the unitarity corrections 
allows to study the $1/Q^2$ behavior of shadowing 
in this model. In order to keep only the leading-twist
contribution (terms $\propto 1/Q^2$
in the cross section) we ignore the higher-twist
contribution of the $S$-component
to the diffractive cross section\footnote{Concretely,
we ignore the
$S$-component in (17) of \cite{model2}, and in (20) of \cite{model2}
we set the exponential containing $\chi_S$ to 1.}.
The results are given in Fig. 10. One can see that neglecting
these terms introduces only a small difference. The fact 
that nuclear shadowing corrections are predominantly leading-twist is not
unexpected, as the diffractive cross section is also leading-twist for the
relevant kinematical region
(indeed,
in the model of \cite{model2} 
the $S$-component diffraction is almost negligible for small $Q^2$
and/or large $M^2$). This is also
seen in the fact that the ratio of diffractive
to inclusive cross sections does not show any strong $Q^2$-dependence 
for large $M^2$ \cite{h1prel}.
Here a comment is in order: in \cite{model2} a parameter $s_0$ is introduced
in $x$ and $\beta$
to control the limit $Q^2\to 0$, so that all the equations are written for
$\bar x=x+s_0/(W^2+Q^2)$, $\bar \beta=\beta+s_0/(M^2+Q^2)$. These terms
could mimic higher-twist corrections. In Fig. 10 we check that the effect
of varying this parameter\footnote{And setting $c=0$ in Eq. (27) of
\cite{model2}.} from the original $s_0=0.79$ GeV$^2$
to $s_0=0.2$ GeV$^2$
is also very 
small. So, we can conclude that the contribution from
higher-twist terms
to the shadowing of $F_2$ is small. 
In contrast, 
in \cite{fgms} a large higher-twist
correction for the shadowing is claimed. The approach in this reference is 
very similar to ours: the authors also compute shadowing from the diffractive
cross section, but using the H1 parametrization
\cite{h1diff}. The shadowing obtained in this way for $Q^2_0=4$ GeV$^2$ is then
employed as initial condition for DGLAP evolution, taking
the shadowing for valence quarks from \cite{gnucl1}.
When evolved downwards to $Q^2=3$ GeV$^2$ a disagreement is found with
experimental data on the ratio Ca over D. This disagreement is attributed
to higher-twist contributions.

\section{Multiplicity reduction in nucleus-nucleus collisions} \label{mulred}

In this framework it is also possible to study the reduction of multiplicities
in nucleus-nucleus collisions \cite{schnuc,hip,aldol}.
We will denote the rapidity of the produced system in the center of mass frame
by $y^*$.
Shadowing as a function of the rapidity of the produced particle
can be computed taking into account the general relation with the
diffractive variables:
\beq
\label{eq17}
y=\ln\left(\frac{1}{x_P}\right)=\ln\left(\frac{s}{M^2}\right).
\eeq

Then the factor for reduction of multiplicities at fixed impact parameter $b$
is \cite{schnuc,hip,aldol}
\beq
\label{eq18}
R_{AB}(b)=\frac{\int d^2s\  R_A(\vec{s})R_B(\vec{b}-\vec{s})}{T_{AB}(b)}\ .
\eeq
$R_{A(B)}(b)$ is given by the r.h.s.
of (\ref{eq15}) multiplied by $T_{A(B)}(b)$ and
with $f(x,Q^2)$
substituted by $F(s,y^*)$ (see below), and
\beq
\label{eq19}
T_{AB}(b)=\int d^2s\  T_A(\vec{s})T_B(\vec{b}-\vec{s}).
\eeq
(\ref{eq18}) takes into account the summation of Schwimmer's fan-like diagrams
for the projectile and target, which are joined by a single Pomeron
whose cut gives rise to the produced particle (Fig. 11).
It follows from AGK cancellation \cite{agk} that
this is the only
contribution of this type (more complicated
diagrams with lines joining upper and lower parts of the diagram,
cancel).
This provides the justification for the factorized expression (\ref{eq18}),
which is true even if more general rescattering diagrams are taken into account.

The reduction factor as a function of the rapidity of the produced particles
$F(s,y^*)$, can be calculated in several ways. The first one is using
(\ref{eq11}), but with the integration limits
inspired by the parton model for hard processes:
for projectile $A$ (target $B$),
\beq
\label{eq20}
x_{A(B)}=\frac{m_T}{\sqrt{s}}e^{\pm y^*},
\eeq
with $y^*>0$ for the projectile hemisphere and $y^*<0$ for the target one, and
$m_T=\sqrt{m^2+p_T^2}$ the transverse mass of the produced particle.
Substituting in the general relation for $M^2_{max}$, (\ref{eq5}), we get
\beq
\label{eq21}
M_{max}^{2(A(B))}=Q^2\left(\frac{x_{Pmax}}{x_{A(B)}}-1\right)
=Q^2\left(\frac{x_{Pmax}\sqrt{s}}{m_T}e^{\mp y^*}-1\right),
\eeq
while $M^2_{min}$ remains fixed and equal to 0.08 GeV$^2$, and $Q^2=m_T^2$.

On the other hand, we can also compute the reduction factor from the formulas
\cite{hip,aldol}
\beq
\label{eq22}
F(s,y^*)
=4\pi\int_{y_{min}}^{y_{max}}dy\ \frac{1}{\sigma_P(s)}\left.\frac{d\sigma^{PPP}}
{dydt}\right\vert_{t=0}F_A^2(t_{min}),
\eeq
where $\sigma_P(s)$ is the single Pomeron exchange cross section and 
$\frac{d\sigma^{PPP}}{dydt}$ the triple Pomeron cross section.
Using the standard triple Pomeron formula for the latter, we get
\beq
\label{eq23}
\frac{1}{\sigma_P(s)}\left.\frac{d\sigma^{PPP}}{dydt}\right\vert_{t=0}=
C\Delta \exp{(\Delta y)},
\eeq
with $C=\frac
{g_{pp}^{P}(0)r_{PPP}(0)}{4\Delta}$,
$g_{pp}^{P}(0)$ the Pomeron-proton coupling and $r_{PPP}(0)$
the triple Pomeron coupling, both evaluated at $t=0$.
In this case,
the same integration limits used above correspond to:
\beq
\label{eq24}
y^{(A(B))}_{min}=\ln{\left(\frac{s}{M^{2(A(B))}_{max}}\right)},
\eeq
with $M^{2(A(B))}_{max}$ given by (\ref{eq21}), and
\beq
\label{eq25}
y_{max}^{(A(B))}=\frac{1}{2}\ln{\left(\frac{s}{m_T^2}\right)} \mp y^*.
\eeq
In the calculations we have used
[$C=0.31$ fm$^2$,$\Delta=0.13$] taken from \cite{ckmt}
(used in \cite{hip,aldol}).
A value $m_T=0.4$ GeV is employed by default (in \cite{hip,aldol} the nucleon
mass $m_N$ was used). The sensibility of our results
to variations in $m_T$ will be
examined.

In Fig. 12 our results at $y^*=0$ are presented for AuAu at RHIC energies
and for PbPb collisions at the LHC, versus impact parameter. Reductions of
multiplicities at $b=0$ by factors
$\sim 1/2$ for RHIC and $\sim 1/3$ for LHC are found, with a clear increase
of the suppression
with increasing
energy. In Fig. 13 results are presented for AuAu at RHIC and PbPb
at LHC for different $y^*$. Finally, in Fig. 14 the variation with $m_T$ of
the results at $y^*=0$ for AuAu at RHIC is studied. A reduction of the
suppression with increasing $m_T$ is seen, as expected. Let us make two
comments: First,
our results for the reduction factors are very similar to the ones estimated
in \cite{hip,aldol}. It has been shown in \cite{hip,aldol} that, when these
reduction factors are used to correct the results of the Dual Parton Model,
one obtains a good description of the RHIC data on multiplicities and their
evolution with centrality. Thus, our results provide a detailed calculation
of these reduction factors which confirms the estimations in \cite{hip,aldol}.
Second, our results are important in studying particle production
in heavy ion collisions.
In particular, the dependence of the reduction factors on $m_T$ gives the
variation of shadowing corrections with the $p_T$ of the produced
particle\footnote{For reduction factors based on other mechanisms, see
\cite{qm02,arpa}.}.

\section{Conclusions} \label{conclu}

In this paper, we have used the relation
which arises from Reggeon Calculus and the AGK rules,
between the diffractive cross section
measured in DIS on nucleons and the first contribution
(i.e. double scattering) to nuclear shadowing.
The next contributions have been estimated using two different methods for
unitarization. In this way we have obtained a description of nuclear shadowing,
based on the model of \cite{model2} for diffraction, which
agrees with the existing experimental data without any fitted parameter. The
model is designed for the region of $x<0.01$ and $Q^2<10$ GeV$^2$, i.e. small
$x$ and small or moderate $Q^2$.

The same method has been applied in \cite{funnuc,fgms}. In
\cite{funnuc}, a model for diffraction \cite{ckmt} has been used that
takes into account
unitarization effects in an effective manner, so the extrapolation to smaller
$x$ or larger $W^2$ is not so reliable as in the full unitarization program
followed in \cite{model2}; furthermore, the description of diffraction in the
model we use is substantially better due to the inclusion in the fits of new,
more precise experimental data. In \cite{fgms}, a model
for diffraction is used
in order to obtain an initial condition for DGLAP evolution
at $Q_0^2=4$ GeV$^2$, so their leading-twist
description for nuclear shadowing is not valid at
small $Q^2$. On the contrary, we develop a model valid for the full
low $Q^2$ region which does not correspond to any definite twist but
contains contributions from all twist orders.
Nevertheless, it turns out that, as discussed at the
end of Section \ref{numer}, the leading-twist contribution is the
dominant one in our model, which
is in reasonable agreement with the existing experimental data.
Precise data on the $Q^2$-dependence of nuclear structure functions should
disentangle between these two possibilities. The existing data
from NMC \cite{nmc-2}
can be well reproduced within the leading-twist DGLAP evolution
\cite{gnucl1} with an appropriate set of initial conditions.
An extension of our results using DGLAP evolution for large values of $Q^2$
is thus a natural continuation
of our work \cite{novonoso}.

In this framework we have also obtained the factor for multiplicity reduction
in nucleus-nucleus collisions. This factor reaches values $\sim 1/2$ and
$\sim 1/3$ for central AuAu and PbPb collisions at RHIC and LHC respectively.
It is therefore a very important ingredient for the computation of particle
production at these energies which should be taken into account together with
other possible effects.

Comparison among models shows differences of a factor 0.6 for the ratio
of structure functions Pb/nucleon at $x=10^{-5}$ and
$Q^2=3$ GeV$^2$. These differences have a large impact in the computation
of particle production in nuclear collisions at energies of RHIC and LHC. They
should be testable in future lepton-ion
colliders \cite{eic}.

To conclude, the method which we have followed offers a natural link between the
measurements of nucleon diffractive structure functions and nuclear shadowing,
and between the latter and the suppression of particle production in nuclear
collisions. In this way the study of Low $x$ Physics at HERA is linked to
that of nuclear structure functions at future lepton-ion colliders and
with Heavy Ion Physics at RHIC and LHC \cite{Mueller:2002kw}.

\vskip 1cm

\noi {\bf Acknowledgments:}
The authors express their gratitude to B. Badelek, E. G. Ferreiro,
B. Z. Kopeliovich, J. Raufeisen and M.
Strikman for useful comments and discussions. They also thank M. Lublinsky for
discussions and for providing the results \cite{jochen}, and C. Pajares for
a reading of the manuscript.
A. C., A. B. K. and C. A. S. acknowledge
financial support by grant INTAS 00-00366, and A. B. K. by grants
RFBR 00-15-96786, RFBR 01-02-17383 and DFG 436 RUS 113/721/0-1. N. A.
thanks the Institute for Nuclear Theory at the University of Washington for
its hospitality and the US Department of Energy for partial support during
the completion of this work.
J. L.-A. thanks CERN Theory Division for kind hospitality, and Universidad de
C\'ordoba and Ministerio de Cultura y Deporte of Spain (grant AP2001-3333)
for financial
support. C. A. S. is supported by a Marie Curie
Fellowship of the European Community programme TMR (Training and
Mobility of Researchers), under the contract number
HPMF-CT-2000-01025.

\section*{Figure captions:}
 
\noi {\bf Fig. 1.} Diagram showing diffractive DIS with the corresponding
kinematical variables in the infinite momentum frame (left) and its
equivalence in the rest frame of the nucleon (right).
 
\noi {\bf Fig. 2.} Diagram showing the equivalence between diffractive DIS
and two exchanged amplitudes with a cut between the amplitudes.
 
\noi {\bf Fig. 3.} Results of the model using Schwimmer (solid lines) and
eikonal
(dashed lines) unitarization compared with experimental data versus $x$,
for the ratios
C/D, Ca/D,
Pb/D \cite{e665-2} and Xe/D \cite{e665-1} (filled circles correspond to the
analysis with hadron requirement and open circles to that with
electromagnetic cuts, see the experimental paper for more details).
 
\noi {\bf Fig. 4.} Results of the model using Schwimmer (open circles) and
eikonal
(open triangles) unitarization compared with experimental data versus $A$,
for the ratios
Be/C, Al/C, Ca/C, Fe/C, Sn/C and Pb/C \cite{nmc-1} at two fixed
values of $x$.
 
\noi {\bf Fig. 5.} Id. to Fig. 3 but for the ratios He/D, C/D and Ca/D
\cite{nmc-3}.
 
\noi {\bf Fig. 6.} Results of the model using Schwimmer (solid lines) and
eikonal
(dashed lines) unitarization compared with experimental data versus $Q^2$,
for the ratio
Sn/C \cite{nmc-2} at two fixed
values of $x$.
 
\noi {\bf Fig. 7.} Comparison of the results of our model
using Schwimmer (solid lines) and
eikonal
(dashed lines) unitarization for the ratio Pb/nucleon
with other models, versus $x$ at fixed $Q^2=3$ GeV$^2$. HKM are the results from
\cite{gnucl3}, Sarcevic from \cite{ina}, Bartels from \cite{jochen},
Frankfurt from \cite{fgms} (at
$Q^2=4$ GeV$^2$),
Armesto from \cite{meu} and EKS98 from \cite{gnucl1}.
 
\noi {\bf Fig. 8.} Results of the model using Schwimmer (solid lines) and
eikonal
(dashed lines) unitarization for the ratios D/nucleon, He/nucleon,
Li/nucleon, C/nucleon, Ca/nucleon, Sn/nucleon and Pb/nucleon versus $x$ at
$Q^2=0.5$, 2 and 5 GeV$^2$.
 
\noi {\bf Fig. 9.} Results of the model for $Q^2=0$
using Schwimmer (solid lines) and
eikonal
(dashed lines) unitarization for the ratios D/nucleon, He/nucleon,
Li/nucleon, C/nucleon, Ca/nucleon, Sn/nucleon and Pb/nucleon
(upper plot), and for different impact parameters $b$ for the ratios C/nucleon
(plot in the middle) and Pb/nucleon (lower plot), versus $W^2$.
 
\noi {\bf Fig. 10.} Results for $Q^2=0.5$ (upper plot) and 5
(lower plot) GeV$^2$ using Schwimmer unitarization for the ratio Pb/nucleon
versus $x$,
of the model without modifications (solid lines), without the higher-twist
contribution in the short-distance component (dashed lines), and without the
higher-twist 
contribution in the short-distance component plus some modification in
parameters (dotted lines) to check the sensibility of the results, see text.
 
\noi {\bf Fig. 11.} Diagram showing the contribution to particle
production in the central region in AB collisions.
 
\noi {\bf Fig. 12.} Results of the model for the multiplicity reduction factor
versus impact parameter $b$ at $y^*=0$, for AuAu collisions at $\sqrt{s}=19$,
130 and
200 GeV per nucleon, and for PbPb collisions at $\sqrt{s}=5500$ GeV
per nucleon,
in the parton model-like realization (solid lines) and for
[$C=0.31$ fm$^2$,$\Delta=0.13$]
(dashed lines).

\noi {\bf Fig. 13.} Id. to Fig. 12 but for AuAu collisions at $\sqrt{s}=200$
GeV per nucleon and for PbPb collisions at $\sqrt{s}=5500$ GeV per nucleon,
for $y^*=1$, 2 and 3.
 
\noi {\bf Fig. 14.} Results of the model for the multiplicity reduction factor
versus impact parameter $b$ at $y^*=0$, for AuAu collisions at $\sqrt{s}=200$
GeV per nucleon, in the parton model-like realization (upper plot) and for
[$C=0.31$ fm$^2$,$\Delta=0.13$]
(lower plot).
In each plot, lines from bottom to top correspond to $m_T^2=0.16$, 1, 2, 3, 4
and 5 GeV$^2$.
 
\newpage
\centerline{\bf \Large Figures:}
 
\vskip 3cm
 
\begin{center}
\epsfig{file=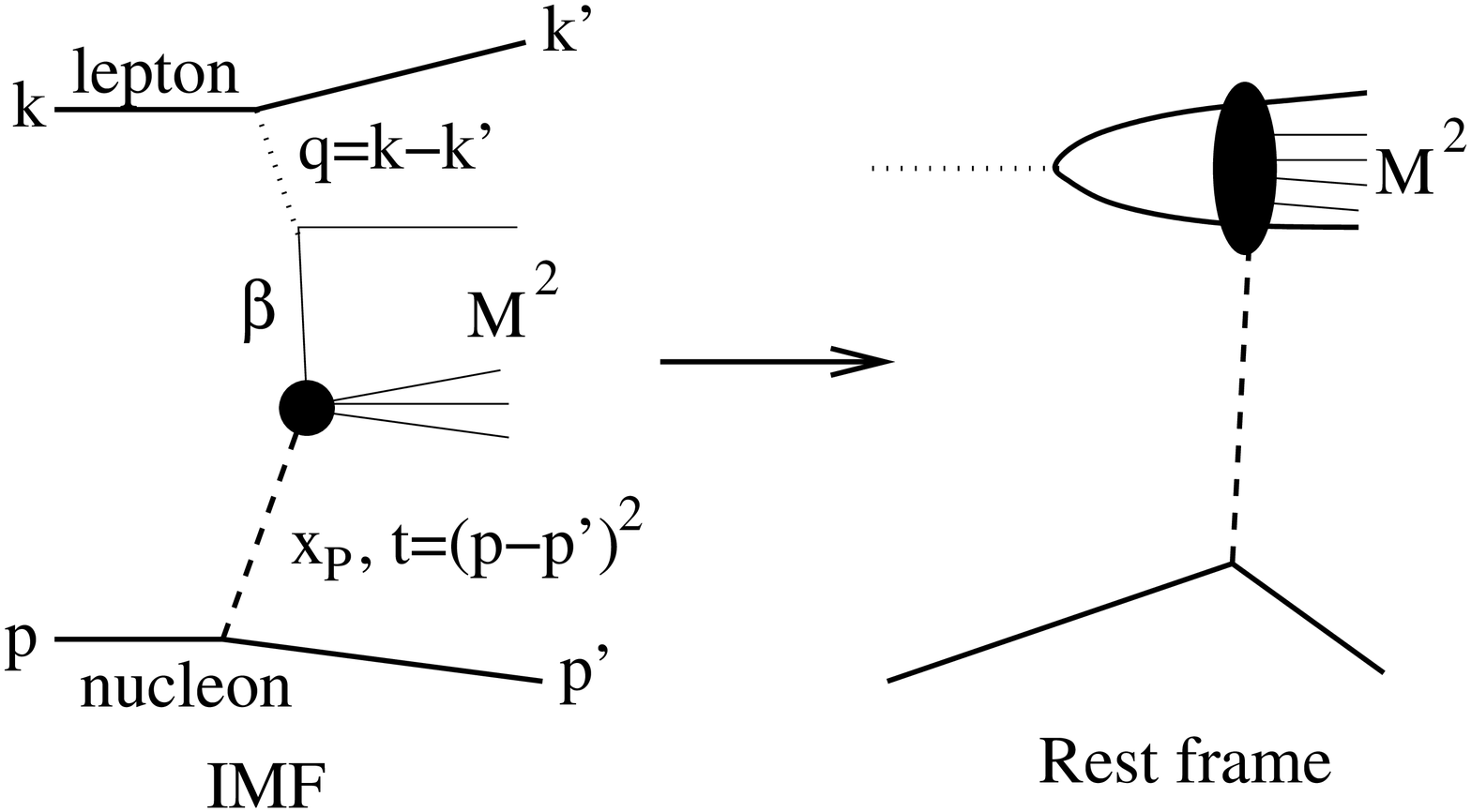,width=15.5cm}
\vskip 1cm
{\bf \large Fig. 1}
\end{center}

\newpage
\centerline{  }
 
\vskip 3cm
 
\begin{center}
\epsfig{file=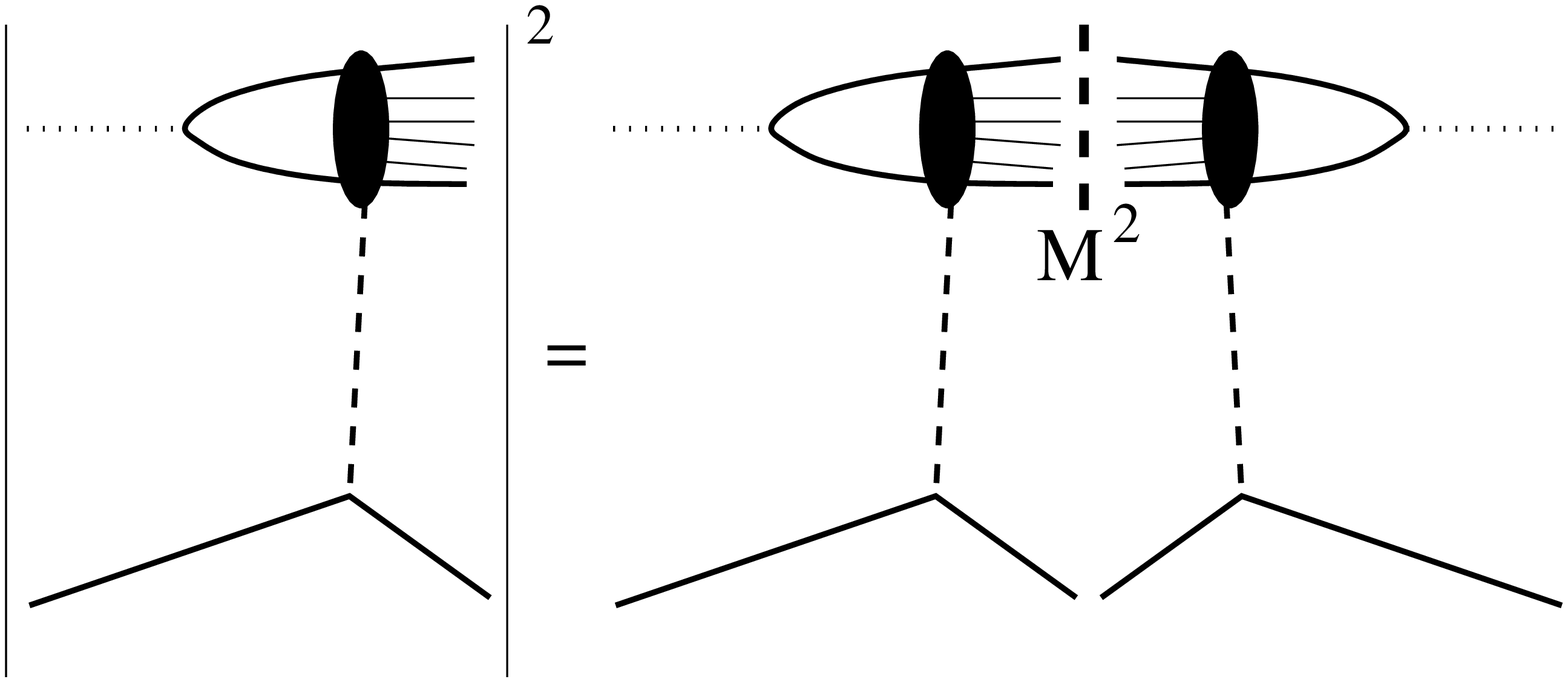,width=15.5cm}
\vskip 1cm
{\bf \large Fig. 2}
\end{center}

\newpage
\centerline{  }
 
\vskip 3cm
 
\begin{center}
\epsfig{file=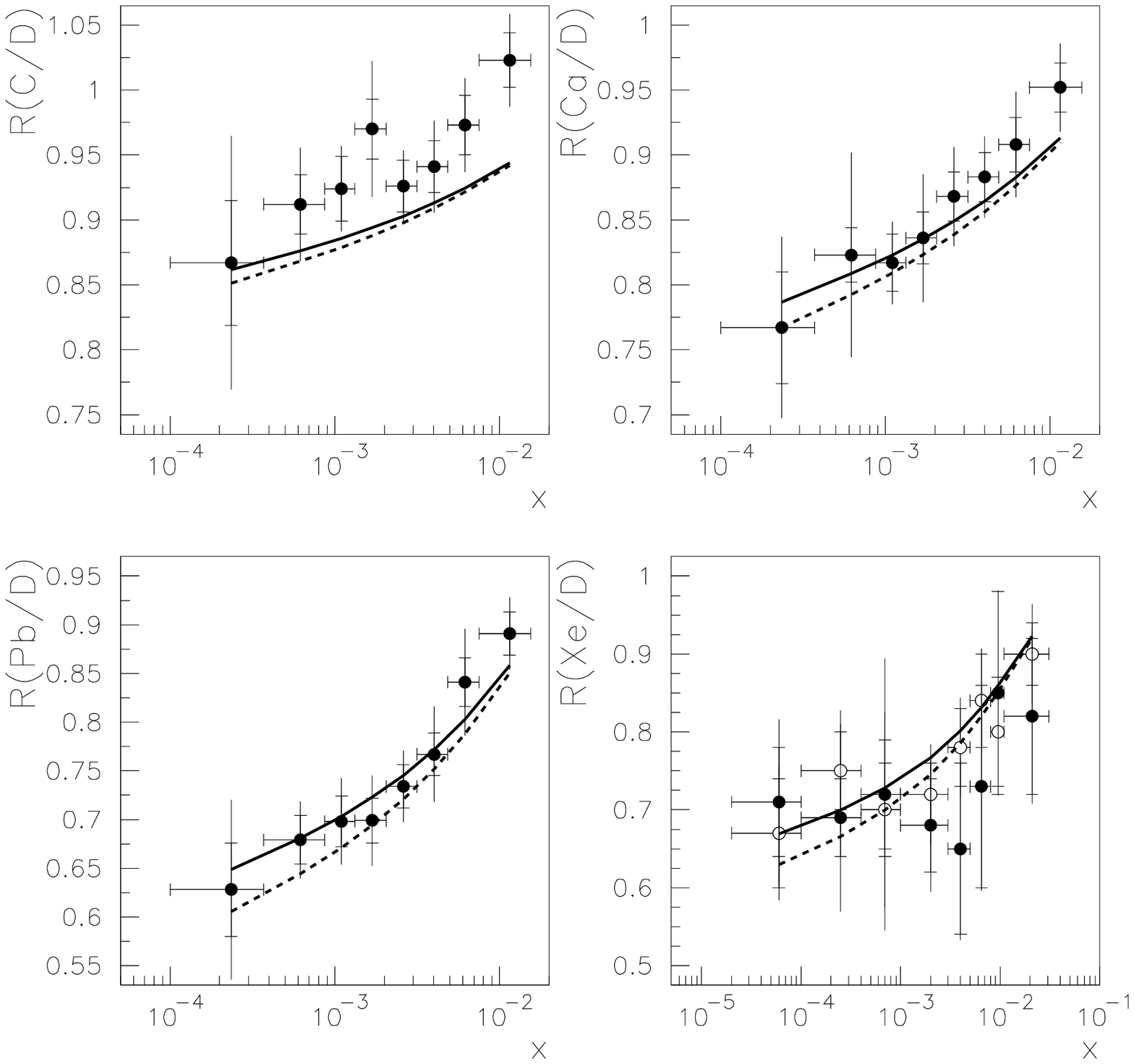,width=15.5cm}
\vskip 1cm
{\bf \large Fig. 3}
\end{center}

\newpage
\centerline{  }
 
\vskip 3cm
 
\begin{center}
\epsfig{file=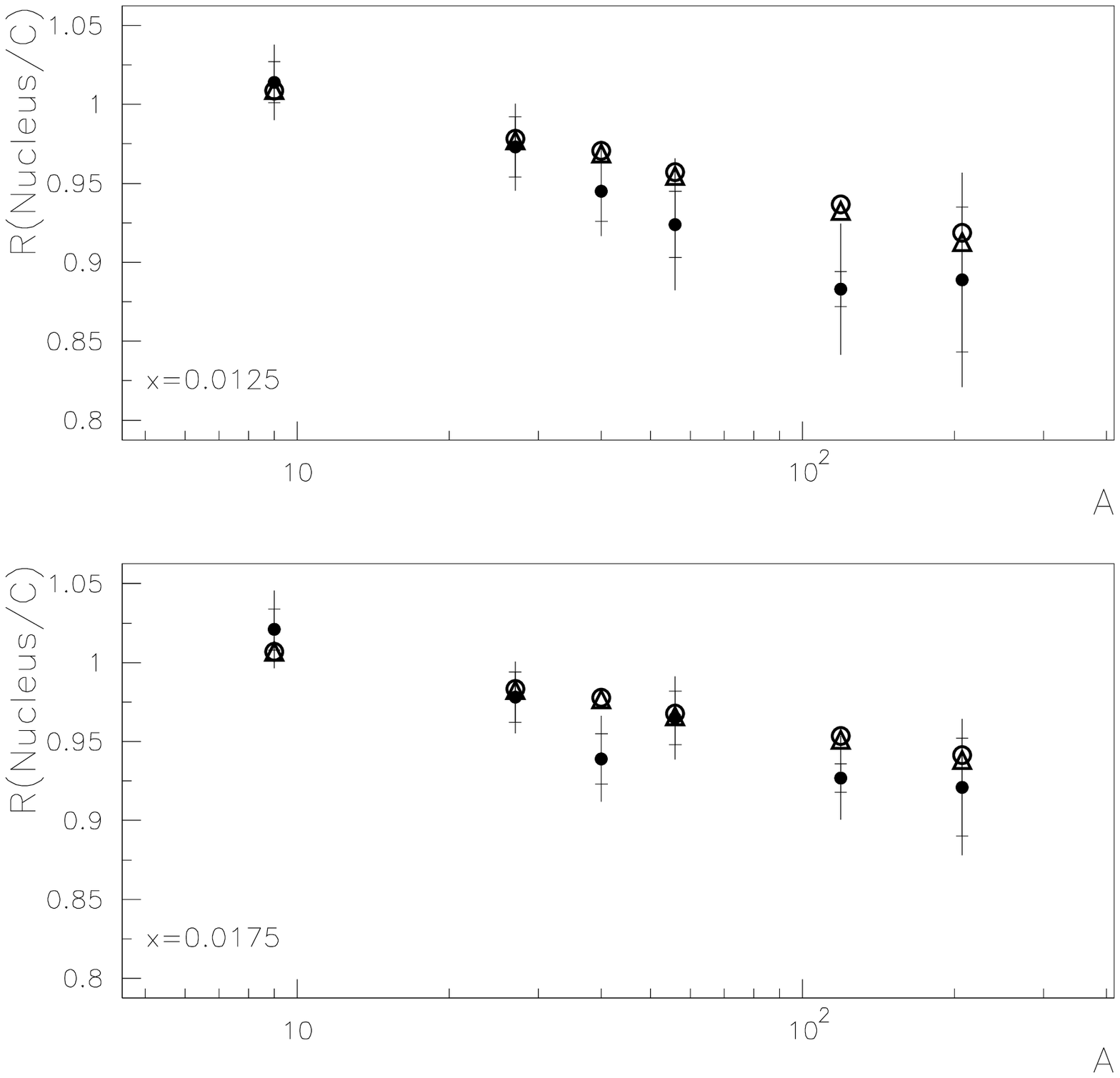,width=15.5cm}
\vskip 1cm
{\bf \large Fig. 4}
\end{center}

\newpage
\centerline{  }
 
\vskip 3cm
 
\begin{center}
\epsfig{file=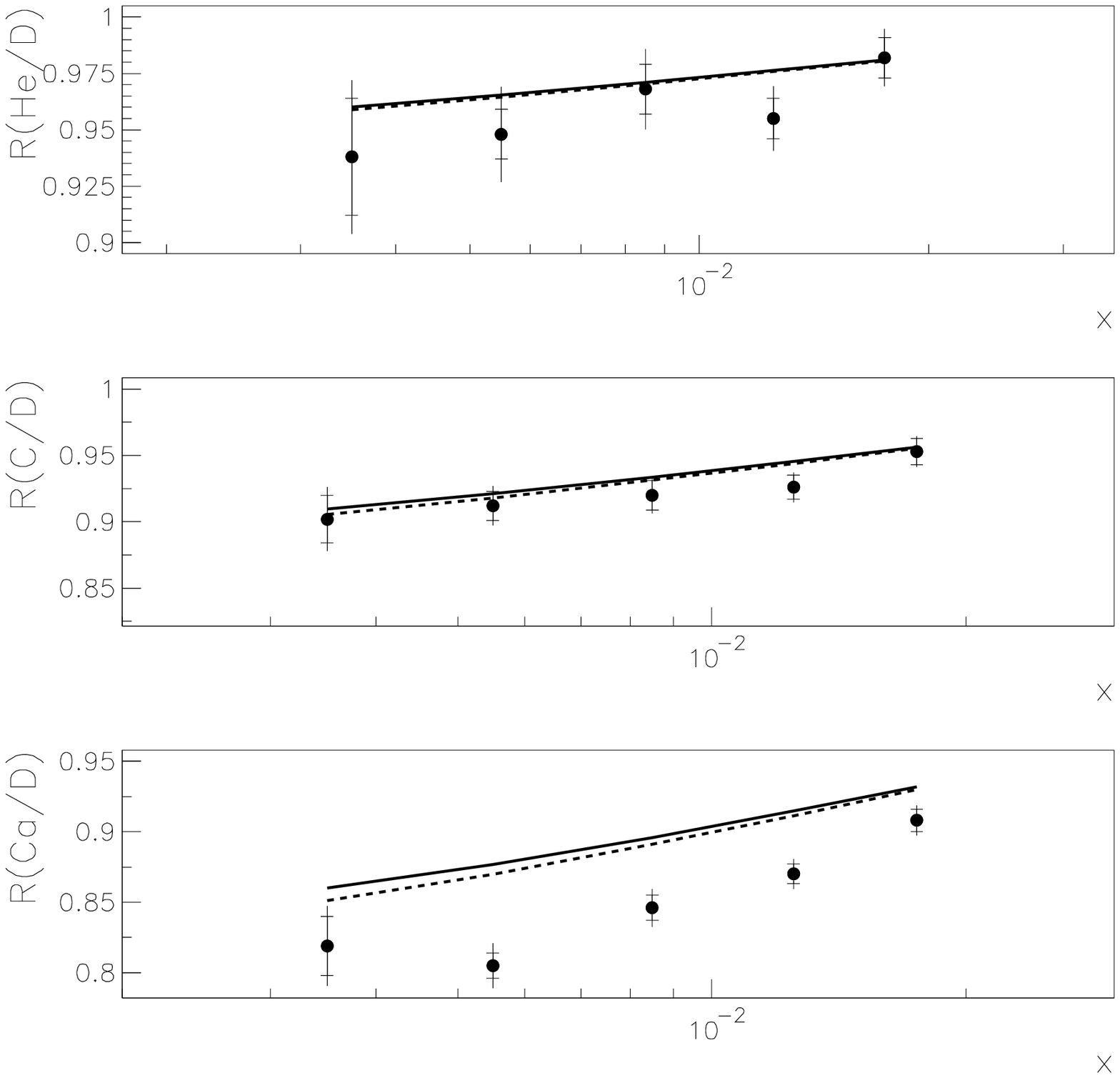,width=15.5cm}
\vskip 1cm
{\bf \large Fig. 5}
\end{center}

\newpage
\centerline{  }
 
\vskip 3cm
 
\begin{center}
\epsfig{file=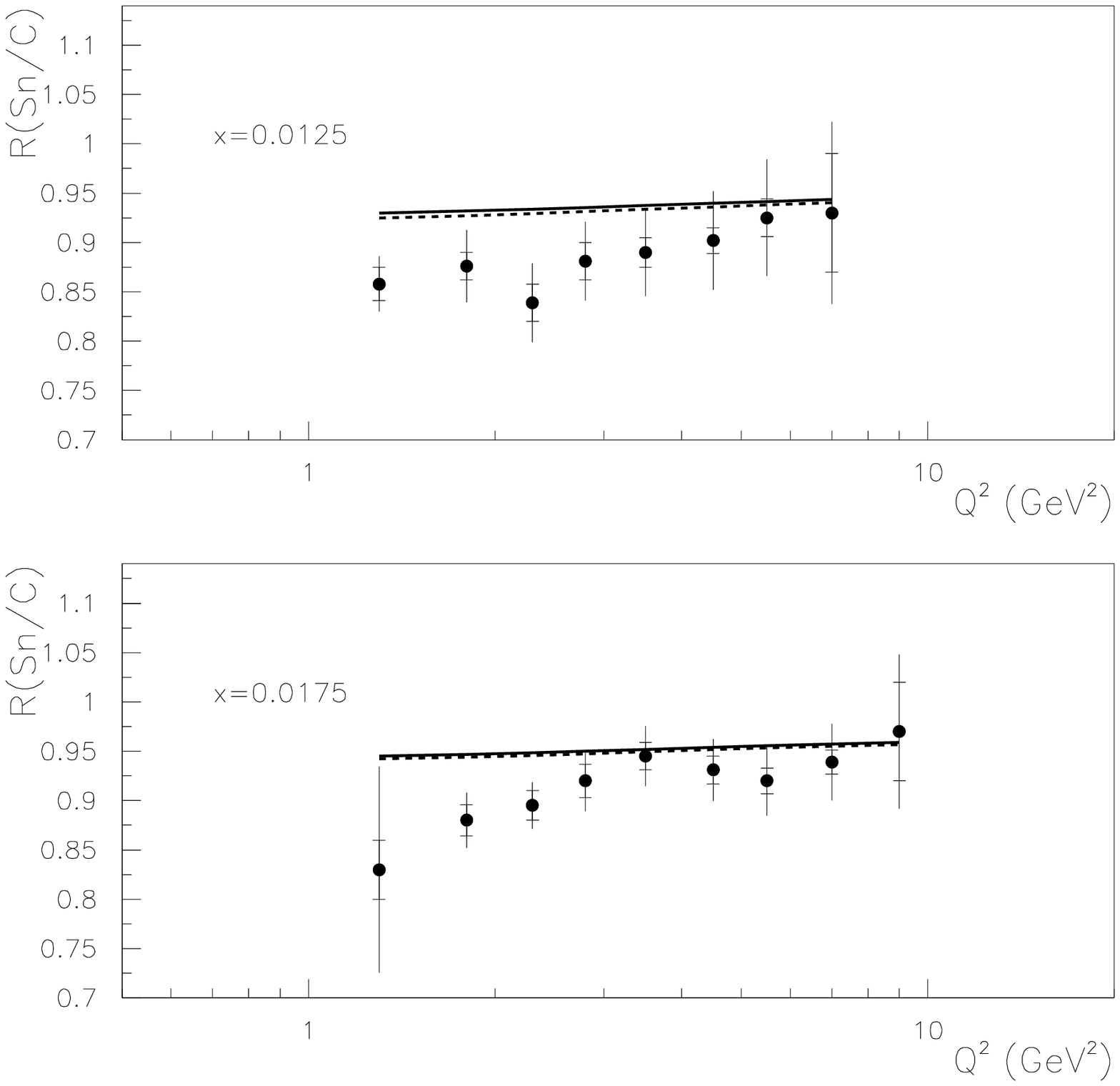,width=15.5cm}
\vskip 1cm
{\bf \large Fig. 6}
\end{center}

\newpage
\centerline{  }
 
\vskip 3cm
 
\begin{center}
\epsfig{file=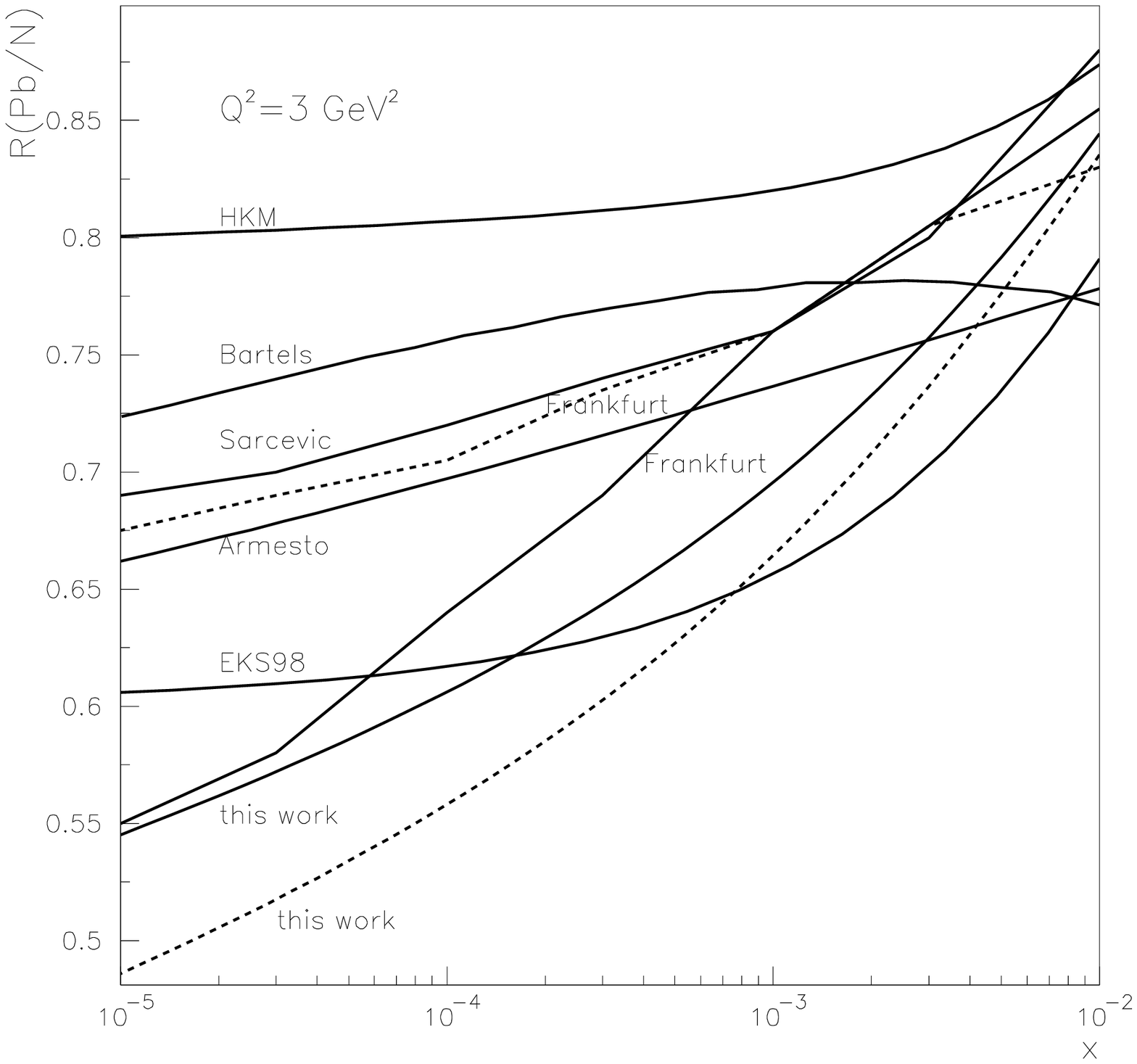,width=15.5cm}
\vskip 1cm
{\bf \large Fig. 7}
\end{center}

\newpage
\centerline{  }
 
\vskip 3cm
 
\begin{center}
\epsfig{file=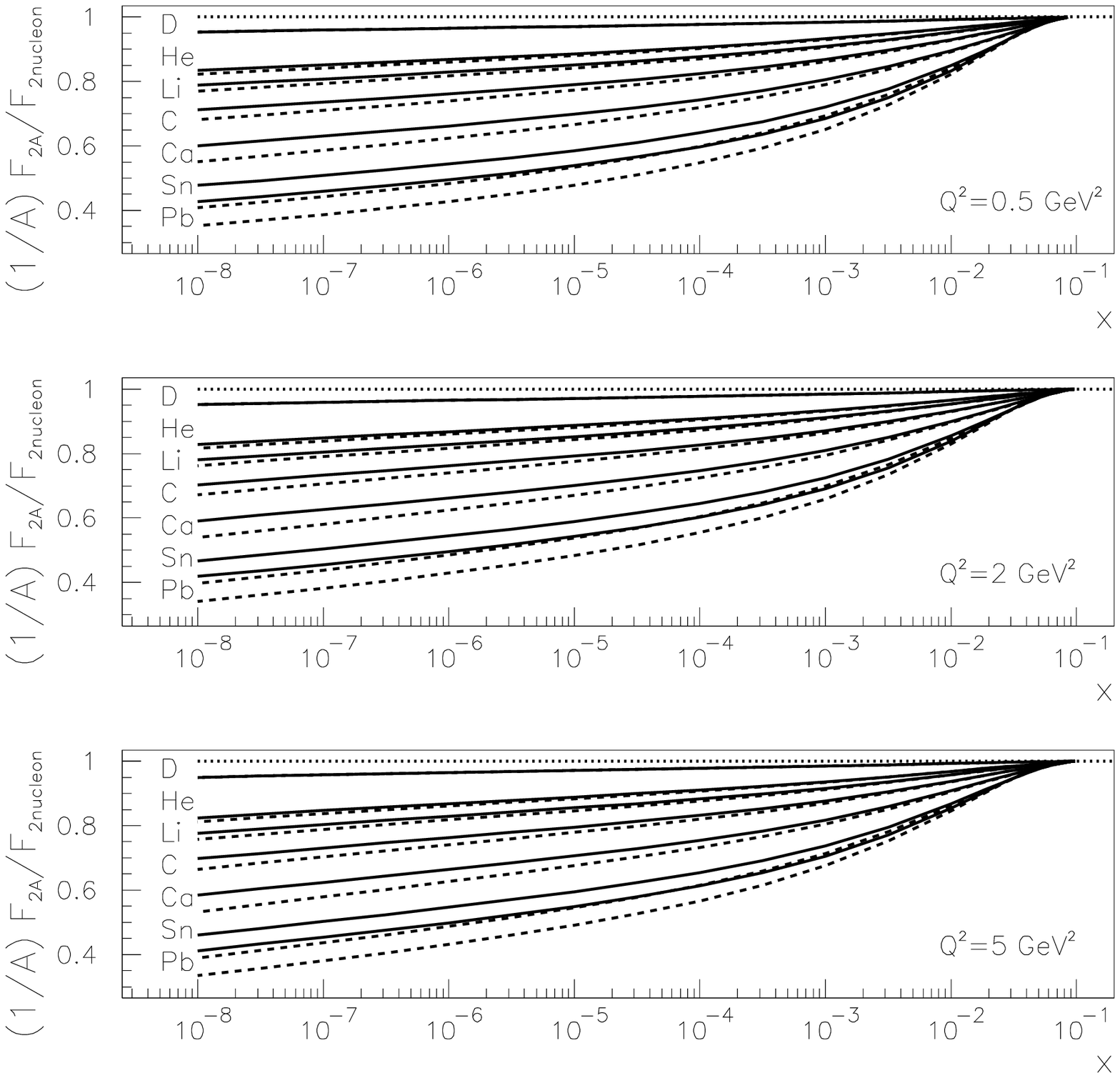,width=15.5cm}
\vskip 1cm
{\bf \large Fig. 8}
\end{center}

\newpage
\centerline{  }
 
\vskip 3cm
 
\begin{center}
\epsfig{file=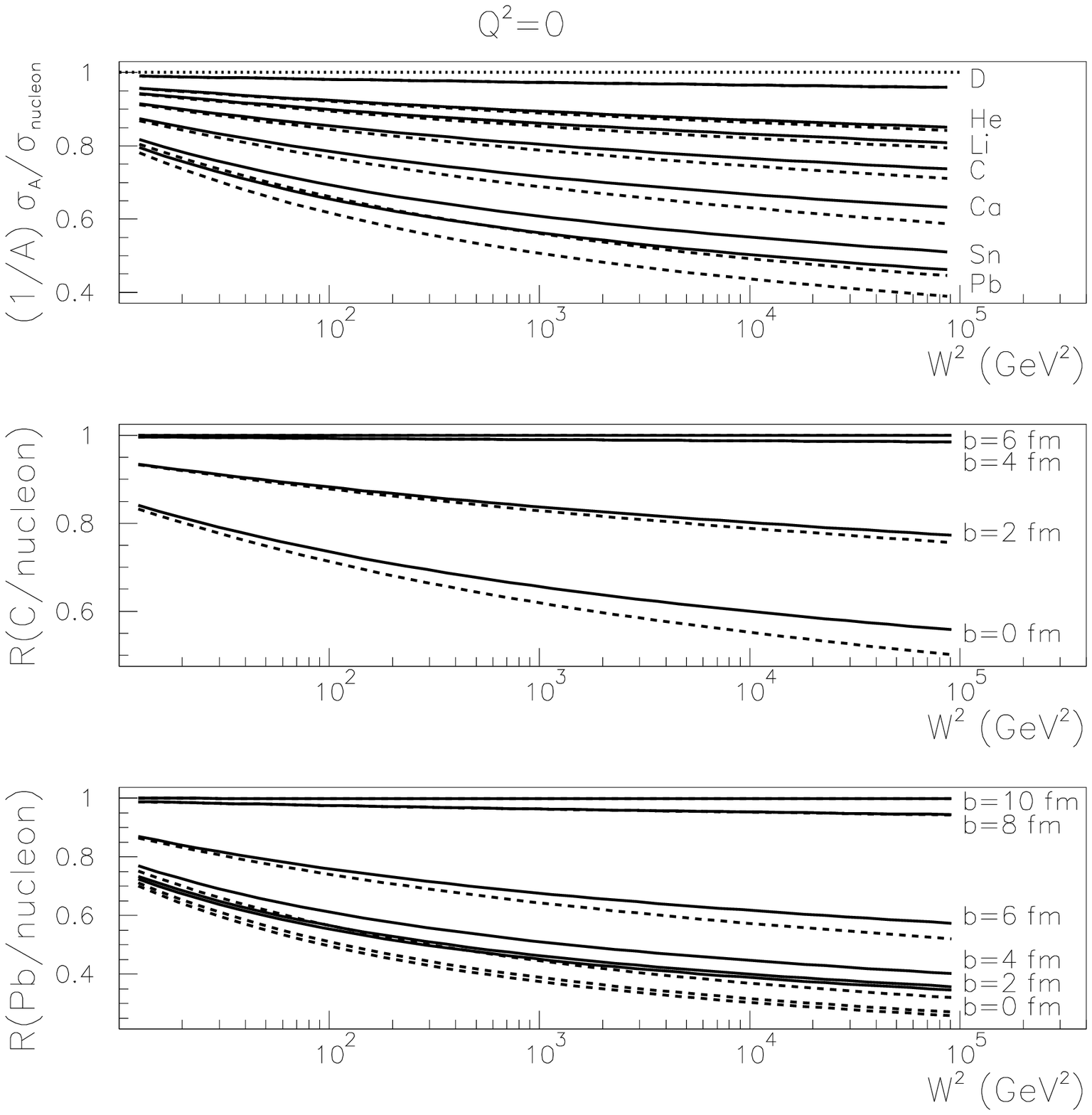,width=15.5cm}
\vskip 1cm
{\bf \large Fig. 9}
\end{center}

\newpage
\centerline{  }
 
\vskip 3cm
 
\begin{center}
\epsfig{file=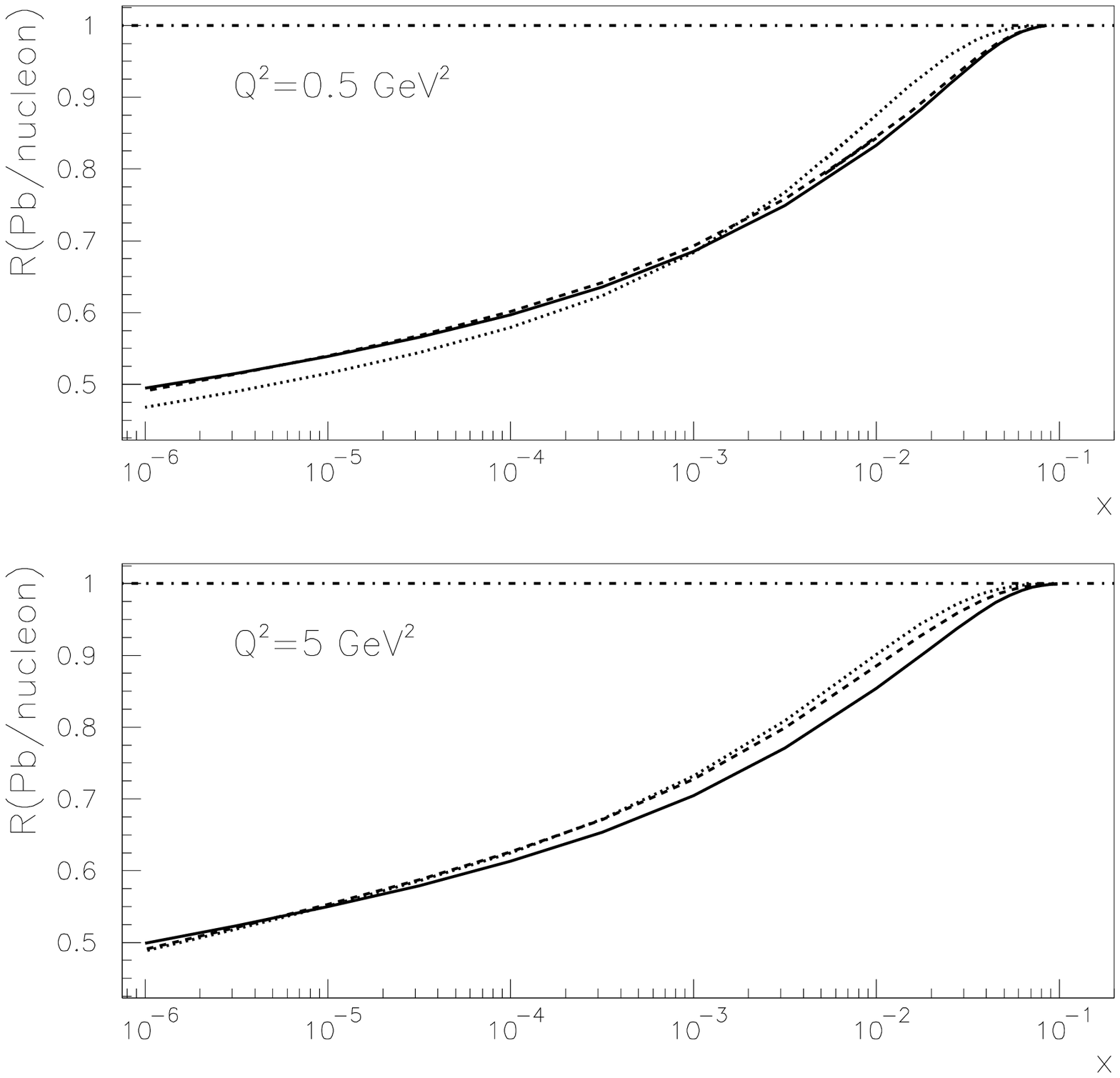,width=15.5cm}
\vskip 1cm
{\bf \large Fig. 10}
\end{center}

\newpage
\centerline{  }
 
\vskip 3cm
 
\begin{center}
\epsfig{file=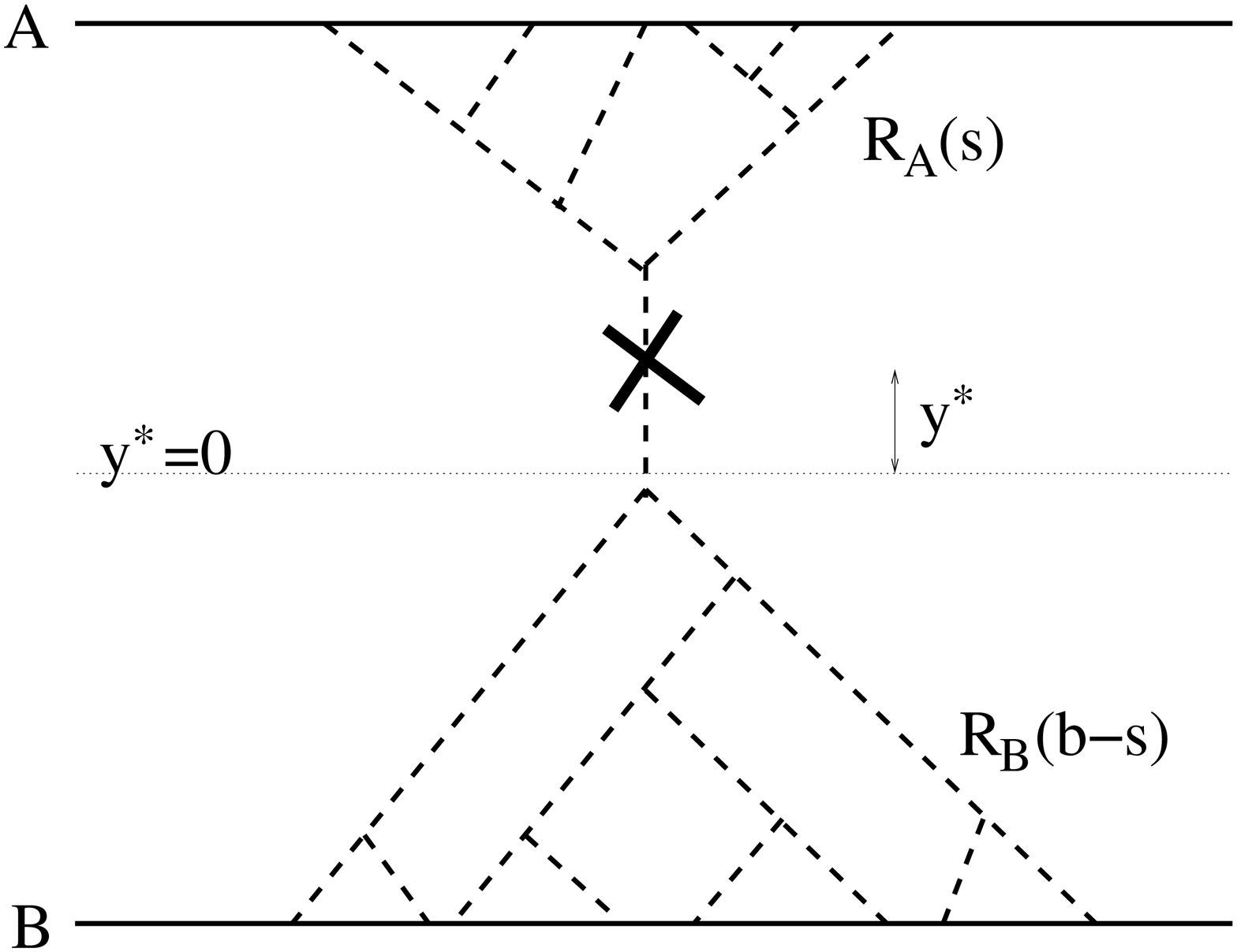,width=15.5cm}
\vskip 1cm
{\bf \large Fig. 11}
\end{center}

\newpage
\centerline{  }
 
\vskip 3cm
 
\begin{center}
\epsfig{file=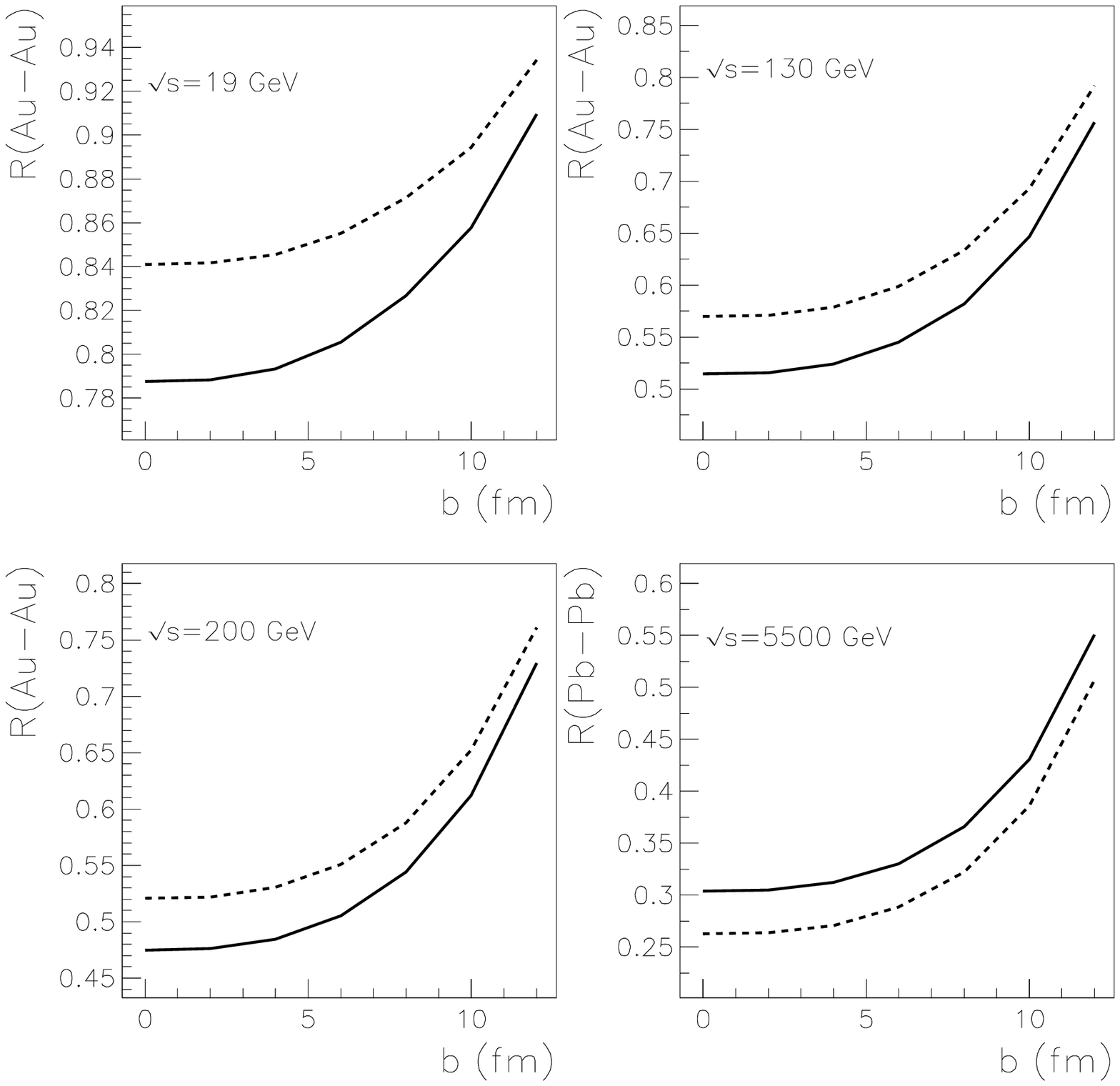,width=15.5cm}
\vskip 1cm
{\bf \large Fig. 12}
\end{center}

\newpage
\centerline{  }
 
\vskip 3cm
 
\begin{center}
\epsfig{file=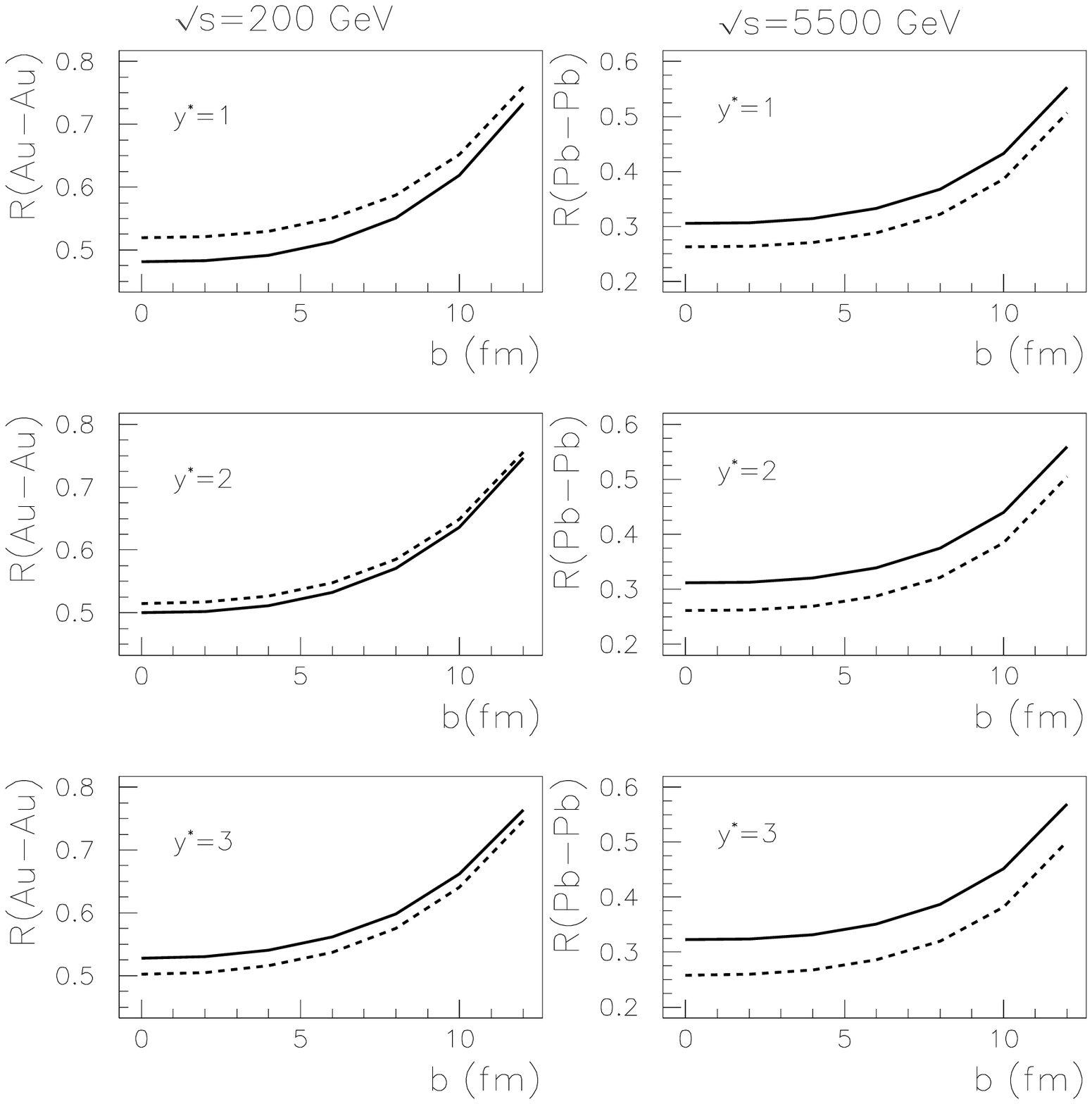,width=15.5cm}
\vskip 1cm
{\bf \large Fig. 13}
\end{center}

\newpage
\centerline{  }
 
\vskip 3cm
 
\begin{center}
\epsfig{file=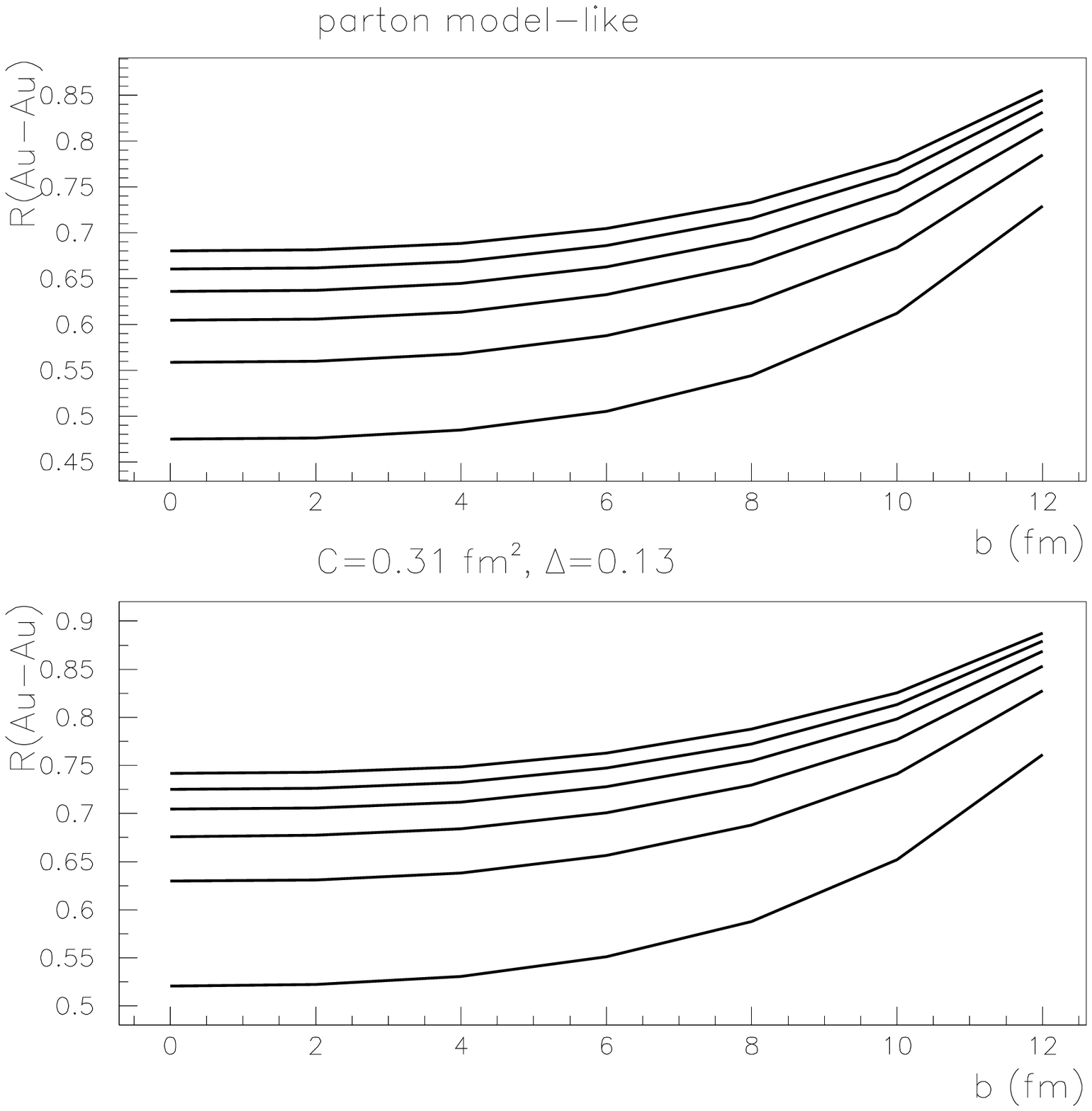,width=15.5cm}
\vskip 1cm
{\bf \large Fig. 14}
\end{center}
 
\end{document}